\documentclass[prl,twocolumn]{revtex4}
\newcommand{\dket}[1]{|  #1 \rangle\!\rangle}
\newcommand{\dbra}[1]{\langle\!\langle #1  |}
\begin{document}

\title{Optimal discrimination of quantum operations}
\author{Massimiliano F. Sacchi}
\affiliation{QUIT, Unit\`a INFM and Dipartimento di Fisica 
``A. Volta'', Universit\`a di Pavia, I-27100 Pavia, Italy.} 
\date{\today}

\begin{abstract}
We address the problem of discriminating with minimal error
probability two given quantum operations. We show that the use of
entangled input states generally improves the discrimination. For
Pauli channels we provide a complete comparison of the optimal
strategies where either entangled or unentangled input states are
used.
\end{abstract}

\maketitle

Quantum nonorthogonality is a basic feature of quantum mechanics that
has deep implications in many areas, as quantum computation and
communication, quantum entanglement, cloning, and
cryptography. Nonorthogonality is strongly related to the concept of
distinguishability, and many measures have been defined to compare
quantum states \cite{distmeas1} and quantum processes
\cite{distmeas2}, according to some experimentally or theoretically
meaningful criteria. Since the pioneering work of Helstrom \cite{hel}
on quantum hypothesis testing, the problem of discriminating
nonorthogonal quantum states has received a lot of attention
\cite{rev12}, with some experimental verifications as well
\cite{exper}. The most popular scenarios are the minimal-error
probability discrimination, where each measurement outcome selects one
of the possible states and the error probability is minimized, and the
optimal unambiguous discrimination \cite{unam}, where unambiguity is
paid by the possibility of getting inconclusive results from the
measurement.  Stimulated by the rapid developments in quantum
information theory, the problem of discrimination has been addressed
also for bipartite quantum states, along with the comparison of global
strategies where unlimited kind of measurements is considered, with
the scenario of LOCC scheme, where only local measurements and
classical communication are allowed \cite{walg}.

The concepts of nonorthogonality and distinguishability can be applied
also to quantum operations, namely all physically allowed
transformations of quantum states. Not very much work, however, has
been devoted to the problem of discriminating general quantum
operations, and major efforts have been directed at the case of
unitary transformations \cite{CPR}. In fact, the most elementary
formulation of the problem can be recast to the evaluation of the norm of
complete boundedness \cite{paulsen}, which is in general a very hard
task.  We recall that such a norm entered the quantum information
field as the diamond norm \cite{diam}, and one of its most relevant
application is found in the problem of quantifying quantum capacities
of quantum information channels \cite{hw}.

In this Letter, we address the problem of discriminating with minimal
error probability two given quantum operations.  After briefly
reviewing the case of quantum states, we formulate the problem for two
quantum operations.  Differently from the case of unitary
transformations \cite{CPR}, we show that entangled input states
generally improve the discrimination. We prove that the use of an
arbitrary maximally entangled state turns out to be always an optimal
input when we are asked to discriminate two quantum operations that
generalize the Pauli channel in any dimension. In the case of qubits,
we give a complete comparison of the strategies where either entangled
or unentangled states are used at the input of the Pauli channels,
thus characterizing the channels where entanglement is really useful
to achieve the ultimate minimal error probability in the
discrimination.

\par In the problem of discrimination two quantum states $\rho _1$ and
$\rho _2$, given with a priori probability $p_1$ and $p_2=1-p_1$,
respectively, one has to look for the two-values POVM $\{\Pi_i \geq 0
\,, \, i=1,2\}$ with $\Pi_1+ \Pi_2=I $ that minimizes the error
probability
\begin{eqnarray}
p_E=p_1 \hbox{Tr}[\rho _1 \Pi_2] + p_2 \hbox{Tr}[\rho_ 2 \Pi_1]\;. 
\end{eqnarray}
We can rewrite
\begin{eqnarray}
p_E &=& p_1 - \hbox{Tr}[(p_1 \rho _1 -p_2 \rho_2 ) \Pi_1]
\nonumber \\&= & 
p_2 + \hbox{Tr}[(p_1 \rho _1 -p_2 \rho_2 ) \Pi_2]
\nonumber \\&= & 
\frac 12 \left \{
1- \hbox{Tr}[(p_1 \rho _1 -p_2 \rho_2 ) (\Pi_1 - \Pi_2)]\right \}
\;, 
\end{eqnarray}
where the third line can be obtained by summing and dividing the two
lines above. The minimal error probability can then be achieved by
taking the orthogonal POVM made by the projectors on the support of
the positive and negative part of the Hermitian operator $p_1 \rho_ 1
-p_2 \rho _2$, and hence one has
\begin{eqnarray}
p_E= \frac 12 \left (1 -\Vert p_1 \rho_ 1 -p _2 \rho _2 \Vert _1
\right )\;,\label{pest}
\end{eqnarray}
where $\Vert A\Vert _1 $ denotes the trace norm of $A$. 
Equivalent expressions for the trace norm are the following 
\begin{eqnarray}
\Vert A \Vert _1= \hbox{Tr}\sqrt{A^\dag A}=\max
  _{U}|\hbox{Tr}[UA]|= \sum _i s_i (A)\;, 
\end{eqnarray}
where the maximum is taken over all unitary operators, and $\{s_i
(A)\}$ denote the singular values of $A$. In the case of
Eq. (\ref{pest}), since the operator inside the norm is Hermitian,
the singular values just corresponds to the absolute value of the
eigenvalues.  

\par The problem of optimally discriminating two quantum operations
${\cal E}_1$ and ${\cal E}_2$ can be reformulated into the problem of
finding in the input Hilbert space $\cal H$ the state $\rho $ such
that the error probability in the discrimination of the output states
${\cal E}_1 (\rho )$ and ${\cal E}_2(\rho )$ is minimal.  We are
interested in the possibility of exploiting entanglement in order to
increase the distinguishability of the output states. In this case the
output states to be discriminated will be of the form $({\cal
E}_1\otimes {\cal I}_{\cal K} ) \rho $ and $({\cal E}_2\otimes {\cal
I}_{\cal K}) \rho $, where the input $\rho $ is generally a bipartite
state of ${\cal H}\otimes {\cal K}$, and the quantum operations act
just on the first party whereas the identity map ${\cal I}={\cal
I}_{\cal K}$ acts on the second.

\par In the following we will denote with $p'_E$ the minimal error
probability when a strategy with unentangled input is adopted.  Hence,
without the use of entanglement the minimal error probability is given
by
\begin{eqnarray}
p'_E 
=\frac 12 \left (1- \max _{\rho \in {\cal H}}\Vert p_1 {\cal E}_1
(\rho )- p_2{\cal E}_2(\rho )\Vert  _1\right )
\;,\label{peno}
\end{eqnarray}
whereas, by allowing the use of entangled input states, one has 
\begin{eqnarray}
p_E =\frac 12 \left (1- \max _{\rho \in {\cal H}\otimes {\cal K}}
\Vert p_1 ({\cal E}_1 \otimes {\cal I})
\rho - p_2 ({\cal E}_2\otimes {\cal I})\rho \Vert  _1\right )
\;.\label{pesi}
\end{eqnarray}
The maximum of the trace norm in Eq. (\ref{pesi}) is equivalent to the
norm of complete boundedness \cite{paulsen}, and in fact for
finite-dimensional Hilbert space one can just consider ${\cal K}={\cal
H}$ \cite{paulsen,diam}.  \par From the linearity of quantum
operations, the following property of the trace norm \cite{bhatia}
\begin{eqnarray} \Vert a A +(1-a) B \Vert _1 \leq a \Vert A \Vert
_1 + (1-a) \Vert B \Vert _1 \;  
\end{eqnarray}
with $0\leq a \leq 1$, and the convexity of the set of states, it
follows that in both Eqs. (\ref{peno}) and (\ref{pesi}) the maximum is
achieved by pure states.

\par The use of entanglement generally improves the discrimination,
and such an improvement can be very remarkable when increasing the
dimension of the Hilbert space. Consider for example the 
situation where one has to 
discriminate between the identity map and the completely depolarizing
map, with $\hbox{dim}({\cal H})=d$.  One has
\begin{eqnarray}
&&{\cal E}_1 (|\psi \rangle \langle \psi |)= |\psi \rangle \langle
\psi | \qquad |\psi \rangle \in {\cal H}\;, \nonumber 
\\& &
       {\cal E}_2 (|\psi \rangle \langle \psi |)= \frac I d \qquad
       |\psi \rangle \in {\cal H}\;, \nonumber 
\\& & ({\cal E}_1
\otimes {\cal I}) (|\psi \rangle \langle \psi |)= |\psi \rangle \langle \psi |
\qquad |\psi \rangle \in {\cal H}\otimes {\cal H}\;, \nonumber 
\\& & ({\cal E}_2
       \otimes {\cal I}) (|\psi \rangle \langle \psi |)= \frac I d \otimes
       \hbox{Tr}_1[|\psi \rangle \langle \psi |] \qquad |\psi \rangle
       \in {\cal H}\otimes {\cal H} \;,\nonumber 
\end{eqnarray}
where $I$ denotes the identity matrix, and $\hbox{Tr}_i$ denotes the
partial trace with respect to the $i$th Hilbert space.  

\par Without the use of entanglement, one has
\begin{eqnarray}
p'_E &=&\frac 12 \left(1- \max _{|\psi \rangle}\left \Vert p_1 |\psi
\rangle \langle \psi | -p_2 \frac I d \right \Vert _1\right )
\nonumber \\&= &\frac 12 \left[1-\left( \left| p_1 -\frac {p_2}{
    d}\right |+ p_2 \frac {d-1}{d}\right )\right ] \;, 
\end{eqnarray}
whereas, by considering an input maximally entangled state $|\phi
\rangle $, one obtains the bound
\begin{eqnarray}
p_E &\leq &\frac 12 \left(1- \left\Vert p_1 |\phi \rangle \langle \phi
| - p_2 \frac {I\otimes I} {d^2} \right \Vert _1 \right) \nonumber \\&
= & \frac 12 \left [1-\left( \left| p_1 -\frac {p_2} {d^2}\right |+
p_2 \frac {d^2-1}{d^2}\right )\right ] \;.\label{leq}
\end{eqnarray}
For $p_1=p_2=1/2$, e.g., one has $p'_E=\frac {1}{2d}$ and $p_E \leq
\frac {1}{2d^2}$ [indeed, from what follows, one has equality in
Eq. (\ref{leq}) for any maximally entangled input state].

\par On the other hand, there are situations in which entanglement is
not needed to achieve the ultimate minimal error probability, as in
the case of discrimination between two unitary transformations
\cite{CPR}.

\par Any quantum operation $\cal E$ is a completely positive
map, and hence can be written in the Kraus form
\cite{Kraus}
\begin{eqnarray}
{\cal E}(\rho )= \sum _n K_n \rho K_n ^\dag \;, 
\end{eqnarray}
where $K_n$ are operators on the Hilbert space $\cal H$ of the quantum
system (here on, for simplicity, we consider operations that map states
from $\cal H$ to $\cal H$), and satisfy the completeness relation $\sum
_n K^\dag _n K_n =I$, thus preserving the trace of $\rho $.

\par Using the notation of Ref. \cite{pla} for bipartite vectors
\begin{eqnarray}
\dket{A} &\equiv  &\sum_{n,m} \langle n|A|m \rangle \,|n \rangle \otimes
|m \rangle \nonumber \\&= & 
 A \otimes I \dket{I}= I \otimes A^\tau \dket{I}
\;,\label{nota}
\end{eqnarray}
one can write the evolution under ${\cal E }\otimes {\cal I}$ of a
pure bipartite state $\rho = \dket{\xi }\dbra{\xi }$ (with
$\hbox{Tr}[\rho]=\hbox{Tr}[\xi ^\dag \xi]=1)$ as follows
\begin{eqnarray}
({\cal E }\otimes {\cal I}) \dket{\xi }\dbra{\xi }=
(I \otimes \xi ^\tau )\,\sum _n \dket  {K_n}\dbra{K_n}  \,(I \otimes \xi ^*)\;, 
\end{eqnarray}
where $\tau $ and $*$ denote transposition and complex conjugation on
the basis chosen in Eq. (\ref{nota}).  Then, the minimal error
probability in Eq. (\ref{pesi}) rewrites
\begin{eqnarray}
p_E= \frac 12 \left (1 - \max _ {\scriptsize{\hbox{Tr}[\xi ^\dag \xi]=1 }}\left
\Vert I\otimes \xi ^\tau \Delta I \otimes \xi ^* \right \Vert _1\right
) \;,\label{pexi}
\end{eqnarray}
where $\Delta $ is Hermitian, and in terms of the Kraus
operators $\{K_n ^{(1)}\}$ and $\{K_m ^{(2)}\}$ of the quantum
operations is given by 
\begin{eqnarray}
\Delta = p_1 \sum _n \dket {K_n^{(1)}}\dbra{K_n^{(1)}}
 -p_2 \sum _m
  \dket {K_m^{(2)}}\dbra{K_m^{(2)}} \;.\label{del}
\end{eqnarray}
Notice that a maximally entangled state writes in the notation of
Eq. (\ref{nota}) as $\frac{1}{\sqrt d}\dket{U}$, with $U$ unitary and
$d=\hbox{dim}({\cal H})$. From the invariance of the trace norm 
$\Vert U A V \Vert _1=\Vert A \Vert _1 $ 
for arbitrary unitary operators $U$ and $V$ \cite{bhatia}, one obtains the
following upper bound for the minimal error probability
\begin{eqnarray}
p_E \leq \frac 12 \left (1- \frac 1d \Vert \Delta \Vert _1\right
)\;.\label{ub}
\end{eqnarray}
\par Exploiting unitarily invariance and the polar decomposition of
$\xi ^\tau $ as $\xi ^\tau =UP$ with $U$ unitary and $P$ positive, the
maximum in Eq. (\ref{pexi}) can be searched for positive operators $P$
with $\hbox{Tr}[P^2]=1$, namely
\begin{eqnarray}
p_E= \frac 12 \left (1 - \max _ {\scriptsize{P\geq 0 \,,\, \hbox{Tr}[P^2]=1 }}\left
\Vert I\otimes P \Delta I \otimes P \right \Vert _1\right )
\;.\label{pepi}
\end{eqnarray}
This expression is very suitable for numerical evaluation. Moreover,
the rank of $P$ that achieves the maximum gives directly information
about the usefulness of entanglement. There is no need of entanglement
for the optimal discrimination if and only if the maximum in
Eq. (\ref{pexi}) can be achieved by a rank-one operator $P$.  

\par The minimal error probability can be evaluated when the quantum
operations can be realized from the same set of orthogonal unitaries
(namely $\{U_n \}$ with $\hbox{Tr}[U^\dag _m U_n]=d \delta _{n,m}$) as
random unitary transformations \cite{orth}.  In this case one has
\begin{eqnarray}
{\cal E}_i(\rho )= \sum _n q_n ^{(i)} U_n \rho U^\dag _n\;, \qquad
\sum _n q_n^{(i)}=1 \;\label{ds}
\end{eqnarray}
and hence $\Delta = \sum _n r_n \dket{U_n}\dbra {U_n} $, with $r_n
=p_1 q_n^{(1)}-p_2 q_n^{(2)}$.  The operator $\Delta $ is diagonal on
maximally entangled states with eigenvalues $d r_n$, and the bound in
Eq. (\ref{ub}) then writes $p_E\leq \frac 12 \left (1- \sum_n
|r_n|\right)$.  On the other hand, one has
\begin{eqnarray}
&&\max _ {|\psi  \rangle \in {\cal H}\otimes {\cal H}}
\left \Vert \sum _n r_n (U_n \otimes I)
  |\psi \rangle \langle \psi |
 (U^\dag _n \otimes I) \right \Vert _1
  \nonumber \\ & & \leq \sum _n |r_n | 
\max _ {|\psi \rangle  }
\left \Vert (U_n \otimes I)
  |\psi \rangle \langle \psi |
 (U^\dag _n \otimes I) \right \Vert _1 \nonumber \\ & & 
= \sum_n |r_n| \;.\label{20}
\end{eqnarray}
From Eq. (\ref{20}) one has $p_E\geq \frac 12 (1- \sum_n |r_n|)$, and
together with the upper bound (\ref{ub}), one obtains
\begin{eqnarray}
p_E=\frac 12 \left(1- \sum_n |r_n|\right )=\frac 12 \left (1 - \frac
1d \Vert \Delta \Vert _1 \right ) \;. 
\end{eqnarray}
This result implies that in the case of Eq. (\ref{ds}) the minimal
error probability can always be obtained by using an arbitrary maximally
entangled state at the input.  
\par Notice that by dropping the condition of orthogonality of the
$\{U_n \}$, one just obtains the bounds
\begin{eqnarray}
\frac 12 \left(1- \sum_n |r_n| 
\right )
\leq p_E \leq 
\frac 12 \left(1- \frac 1d 
\Vert \Delta \Vert _1 \right ) \;. 
\end{eqnarray}

\par In the following we consider the case of discrimination of two
Pauli channels for qubits, namely
\begin{eqnarray}
{\cal E}^{(i)}(\rho )= \sum_{\alpha =0}^3 q_\alpha ^{(i) }\sigma
  _\alpha \rho \sigma _\alpha \;, 
\end{eqnarray}
where $\{\sigma _0\,, \sigma _1\,,\sigma _2\,,\sigma _3 \}= \{I\,,
\sigma _x\,,\sigma _y\,,\sigma _z\}$ and $\sum _{\alpha =0}^3 q_\alpha
^{(i)} = 1$.  In particular, we are interested to understand when the
entangled-input strategy is really needed to achieve the optimal
discrimination. The positive operator $P$ in Eq. (\ref{pepi}) can be
parameterized on the computational basis as follows
\begin{eqnarray}
P=\left(
\begin{array}{cc}
x & z \\
z^* & y
\end{array}
\right)
\;, 
\end{eqnarray}
with $x,y\geq 0$, $xy\geq |z|^2$, and $x^2 +y^2 +2 |z|^2=1$. The
strategy with unentangled input corresponds to the values range
$x+y=1$ and $|z|=\sqrt{xy}$ such that $\hbox{rank}(P)= 1$. The
operator $\Delta $ is diagonal on the Bell basis and writes $\Delta
=\sum _{\alpha =0}^3 r_\alpha \dket{\sigma _\alpha }\dbra{\sigma
_\alpha } $, where $r_\alpha =p_1 q^{(1)}_\alpha -p_2 q^{(2)}_\alpha
$.  On the ordered basis $\{|00 \rangle \,,|01 \rangle \,,|10 \rangle
\,, |11 \rangle \}$, one has
\begin{eqnarray}
\Delta =\left( 
\begin{array}{cccc}
a & 0 & 0& c\\
0 & b & d& 0\\
0 & d & b& 0\\
c & 0 & 0& a
\end{array}
\right )\;, 
\end{eqnarray}
with $a=r_0 +r_3$, $c=r_0 - r_3$, $b=r_1 +r_2$, and $d=r_1 -r_2$.
The singular values of $\Delta $ are given by
\begin{eqnarray}
s_i (\Delta )=\{|a\pm c|\,,|b\pm d|\}\;, 
\end{eqnarray}
and for the previous derivation we know that the minimal error
probability can be achieved by using a maximally entangled input
state, with 
\begin{eqnarray}
p_E=\frac 12 \left(1- \frac 12 \sum _{i=0}^3 s_i (\Delta )\right
)= 
\frac 12 \left(1- \sum _{i=0}^3 |r_i| \right)
\;. 
\end{eqnarray}
For the strategy with unentangled input one  has 
\begin{eqnarray}
&&p'_E= \frac 12 \left (1 - \max _ {P'}\left \Vert I\otimes P'
  \,\Delta \, I \otimes P' \right \Vert _1\right ) \;;\nonumber \\& &
  P'= \left(
\begin{array}{cc}
x & \sqrt{x(1-x)}e^{i\phi} 
\\
\sqrt{x(1-x)}e^{-i\phi}  & 1-x
\end{array}
\right)\;,\nonumber \\& & 
\qquad 0\leq x\leq 1\;, \qquad 0\leq \phi \leq 2 \pi\;.
\label{pepi2}
\end{eqnarray}
In this case the operator $\Delta '= I\otimes P' \,\Delta \,I \otimes P' $
inside the trace norm is at most rank-two, and its nonvanishing
singular values write
\begin{eqnarray}
&&s_{1,2}(\Delta ')=\frac 12 \left| a+b \pm 
\right. \nonumber \\& & 
\!\!\!\!\!\!\!\!\!\!
\left.
\sqrt{[(a-b)(1-2x)]^2 +
4x(1-x)(c^2+d^2+2cd \cos (2\phi))}\right | \;.\nonumber  
\end{eqnarray}
The maximum can be obtained from comparing the values of 
$s_1(\Delta ')+s_2(\Delta ')$ just for the extreme points $x=0,1$ and 
the stationary points $x=1/2$ and $\phi =k\pi /2$ with $k$ integer,
and one has 
\begin{eqnarray}
p'_E=\frac 12 \left (1 - M \right )
\;, 
\end{eqnarray}
where 
\begin{eqnarray}
&&M=\nonumber \\& & 
\max \left\{ |a|+|b|\,, \frac 12 \left (|a+b +c+d|+|a+b -c- d|\right
)\,,\right. \nonumber \\& & \left. 
\frac 12 (|a+b +c-d|+|a+b -c+d|) \right \}=
\nonumber \\& &
\max \left\{ |r_0+ r_3|+|r_1+r_2|\,, 
|r_0+ r_1|+|r_2+r_3|\,, \right. \nonumber \\& & \left. 
|r_0+ r_2|+|r_1+r_3| \right \}
\;,\label{emme} 
\end{eqnarray}
and the three cases inside the brackets corresponds to using as input
state an eigenstate of $\sigma _z$, $\sigma _x$, and $\sigma _y$,
respectively. From Eq. (\ref{emme}) one can see that entanglement is
not needed as long as $M=\sum _{i=0}^3 |r_i|$, and this happens in
many situations: $i)$ when the determinant $\hbox{det}(\Delta )=0$, namely
at least one of the $\{r_i \}$ vanishes; $ii)$ when $\hbox{det}(\Delta
) > 0$, so that two of the $\{r_i\}$ are strictly positive and the
other are strictly negative. On the other hand, entanglement is crucial to
achieve the ultimate minimal error probability when
$\hbox{det}(\Delta) < 0$. Among these cases, there are striking
examples where the  channels can be perfectly discriminated only by
means of entanglement. This is the case of two channels of the form
\begin{eqnarray}
{\cal E}_1(\rho )=\sum _{\alpha \neq \beta }q_\alpha  \sigma _\alpha
  \rho \sigma _\alpha \;,\qquad 
{\cal E}_2(\rho )=\sigma _\beta  \rho \sigma _\beta 
\;, 
\end{eqnarray}
with $q_\alpha \neq 0$, and arbitrary a priori probability. This
example can be simply understood, since the entangled-input strategy
increases the dimension of the Hilbert space such that the two
possible output states will have orthogonal support.  

\par In conclusion, we considered the problem of discriminating two
quantum operations with minimal error probability and showed that the
use of entangled input states generally improves the discrimination.
We gave a general upper bound to the minimal error probability, and
the exact solution for generalized Pauli channels. In the case of
qubits, we characterized in a simple way the Pauli channels where the
use of entanglement definitely outperforms the scheme with unentangled
input. We hope that our results will stimulate further research on the
discrimination of quantum operations.

\emph{Acknowledgments.} Stimulating discussions with G. M. D'Ariano
are acknowledged. This work has been sponsored by INFM through the
project PRA-2002-CLON.


\begin{thebibliography}{99}
\bibitem{orth}This generalizes Pauli channels in arbitrary
dimension. An example of such a set is operators of the form $X^kZ^l$
where $X$ and $Z$ act on the basis states $|0 \rangle , ...,|d-1
\rangle $ as $X| j \rangle =|j \oplus 1 \rangle $ and $Z |j \rangle
=e^{2\pi i j/d}|j \rangle $, with $\oplus $ denoting addition modulo
$d$. Other examples and general theory can be found in
Ref. \cite{klap}.
\end{thebibliography}
\end{document}